\documentclass[conference]{IEEEtran}

\usepackage[a4paper, left=0.680in, right=0.680in, bottom=1.049in, top=0.71in]{geometry}

\usepackage{ifpdf}
\ifCLASSINFOpdf
\usepackage[pdftex]{graphicx}
\graphicspath{{./Figures/}}
\DeclareGraphicsExtensions{.pdf,.jpeg,.png}
\else
\usepackage[dvips]{graphicx}
\DeclareGraphicsExtensions{.pdf}
\usepackage[caption=false, font=footnotesize]{subfig}
\fi

\usepackage{tabularx, booktabs}
\usepackage{amssymb}
\usepackage{amsthm} 
\usepackage[ruled,vlined,linesnumbered]{algorithm2e}
\usepackage{aligned-overset} 
\usepackage{array}
\usepackage{balance}
\usepackage{cite}
\usepackage{color}
\usepackage{epstopdf}
\usepackage{enumitem} 
\usepackage{mathtools} 
\usepackage{makecell} 
\usepackage{multicol} 
\usepackage{multirow} 
\usepackage[caption=false,font=small]{subfig}
\usepackage{textcomp}
\usepackage{stfloats}
\usepackage{url}
\usepackage{verbatim}
\usepackage{graphicx}
\usepackage{url}
\usepackage{float}
\usepackage{xcolor}
\usepackage{svg}
\definecolor{cyan}{RGB}{0, 102, 204} %
\definecolor{red}{RGB}{204, 0, 0} %
\definecolor{green}{RGB}{5,107,68} %
\definecolor{purple}{RGB}{102, 51, 153} %
\definecolor{blue}{rgb}{0.06, 0.2, 0.65}

\usepackage{hyperref}
\hypersetup{colorlinks=true,linkcolor=blue,urlcolor=blue, citecolor=blue}

\newcommand\blfootnote[1]{%
  \begingroup
  \renewcommand\thefootnote{}%
  \footnote{#1}%
  \addtocounter{footnote}{-1}%
  \endgroup
}

\begin{document}

\title{AI-Open-RAN for Non-Terrestrial Networks}
\author{
    \IEEEauthorblockN{Tri~Nhu~Do}
    \IEEEauthorblockA{Telecom Neural Detection Lab, Department of Electrical Engineering, Polytechnique Montr\'{e}al, Montreal, QC, Canada.}
    \IEEEauthorblockA{ Email: tri-nhu.do@polymtl.ca }
}
\maketitle

\begin{abstract}
In this paper, we propose the concept of AIO-RAN-NTN, a unified all-in-one Radio Access Network (RAN) for Non-Terrestrial Networks (NTNs), built on an open architecture that leverages open interfaces and artificial intelligence (AI)-based functionalities. This approach advances interoperability, flexibility, and intelligence in next-generation telecommunications. First, we provide a concise overview of the state-of-the-art architectures for Open-RAN and AI-RAN, highlighting key network functions and infrastructure elements. Next, we introduce our integrated AIO-RAN-NTN blueprint, emphasizing how internal and air interfaces from AIO-RAN and the 3rd Generation Partnership Project (3GPP) can be applied to emerging environments such as NTNs. To examine the impact of mobility on AIO-RAN, we implement a testbed transmission using the OpenAirInterface platform for a standalone (SA) New Radio (NR) 5G system. We then train an AI model on realistic data to forecast key performance indicators (KPIs). Our experiments demonstrate that the AIO-based SA architecture is sensitive to mobility, even at low speeds, but this limitation can be mitigated through AI-driven KPI forecasting.
\end{abstract}

\begin{IEEEkeywords}
O-RAN, AI-RAN, 3GPP, air interface, AI, LSTM non-terrestrial network (NTN), forecasting KPI
\end{IEEEkeywords}

\vspace{-1em}
\section{Introduction}

The next generation (NG) of telecommunication systems is expected to bring a revolutionary Radio Access Network (RAN) with massive, scalable, low-latency, and ultra-reliable connectivity, enabling seamless integration with current and future terrestrial and non-terrestrial networks (NTNs).\blfootnote{\thanks{This research was supported by NSERC (Canada) under Grant Nos. RGPIN-2025-05010 and DGECR-2025-00215.}}
To support flexible multi-vendor deployments, the 3rd Generation Partnership Project (3GPP) standardized the Central Unit (CU) / Distributed Unit (DU) functional split (Option~2, F1 interface, Technical Specification (TS)~38.47x / F1 Application Protocol (F1AP)) in Release~15 (Rel-15) and evolved it in Rel-17 introducing NTN-specific enhancements~\cite{Agarwal2025,ts38401}.  
Complementing this, the Open Radio Access Network (O-RAN) Alliance and the Artificial Intelligence--Radio Access Network (AI-RAN) Alliance propose open and intelligent RAN architectures that extend modularity, virtualization, and artificial intelligence/machine learning (AI/ML), i.e., driven intelligence into RAN design~\cite{oran-specs}. Their objectives include cloud-native disaggregation, standardized application programming interfaces (APIs), interfaces (e.g., OpenAirInterface~\cite{oai}), and advanced automation through the RAN Intelligent Controller (RIC) (xApps/rApps) and the Service Management and Orchestration (SMO) framework~\cite{oran-specs}.

Specifically, the AI-RAN and O-RAN Alliances extend 3GPP standards with open interfaces, RAN disaggregation, and AI-driven automation~\cite{etsipas103859}.  
AI is becoming an integral part of RAN evolution. 3GPP Rel-18 introduced a standardized AI/ML framework for NG-RAN, with Rel-19 extending its capabilities~\cite{tr38743}. In O-RAN, programmable control planes via the Near-Real-Time (Near-RT) and Non-Real-Time (Non-RT) RICs and the SMO enable practical deployment of AI-driven RAN optimization. Meanwhile, AI-RAN initiatives aim to accelerate progress toward fully AI-native RANs\cite{airan,Agarwal2025}.  

For NTNs, 3GPP Rel-17 introduced New Radio (NR)-based NTN support, enabling the integration of satellites and other non-terrestrial platforms with terrestrial 5G systems. NTN mobility is dominated by moving satellites and spot beams, causing frequent handovers and ping-pong effects, while large and time-varying delay and Doppler shifts complicate synchronization, random access, measurements, and handovers. Rel-17 addressed these with Global Navigation Satellite System (GNSS) / ephemeris-assisted timing advance and frequency pre-compensation; Rel-18 (5G-Advanced) added L1/L2-triggered mobility and enhanced conditional handover; and Rel-19 further improves inter-CU mobility and tracking-area handling. Mobility management therefore remains a central design concern for NG-RAN in NTN environments.

In this paper, we propose Artificial Intelligence Open RAN (AIO-RAN), a unified RAN architecture that integrates the strengths of AI-RAN and O-RAN on top of standardized 3GPP foundations, specifically tailored for NTNs. The proposed architecture consolidates openness, programmability, and AI-native coordination across vendors, operators, and clouds. Our contributions are twofold. First, we detail the AIO-RAN-NTN blueprint, illustrating how specific interfaces and functional blocks can be realized. Second, we examine the impact of mobility using an OpenAirInterface-based standalone NR next-generation NodeB (gNB) testbed and demonstrate that the proposed AI-driven KPI forecasting method can effectively mitigate mobility-induced performance degradation. 
Due to space limitations, the full-size high-resolution figure, testbed dashboard, data visualizations, AI/ML configurations, more illustrations and training/testing details are available in our \textit{code repository} at \href{https://github.com/tnd-lab/AIO-RAN-NTN}{github.com/TND-Lab/AIO-RAN-NTN}

\vspace{-0.5em}
\section{Proposed Blueprint of AIO-RAN-NTN: Architecture and Interfaces}

3GPP NG-RAN provides the standardized foundation, with logical nodes (CU, DU, Radio Unit (RU)) and functional splits (e.g., Option~2) defining the baseline for openness and flexibility~\cite{Alam2025,ts38401} and references therein. Building on this, our AIO-RAN-NTN design emphasizes hardware unit splitting and open fronthaul connectivity as key enablers for deployment in NTNs.  

\vspace{-0.5em}
\subsection{Open and AI-RAN Architecture Splitting}

\subsubsection{3GPP Open RAN Baseline}
The standardized NG-RAN architecture specifies possible split options and has established Option~2 (F1 interface), decoupling CU, DU, and RU functions, enabling multi-vendor implementations supported by open fronthaul using the enhanced Common Public Radio Interface (eCPRI).  

\begin{figure*}[htp]
	\centering
	\includegraphics[width=.7\linewidth]{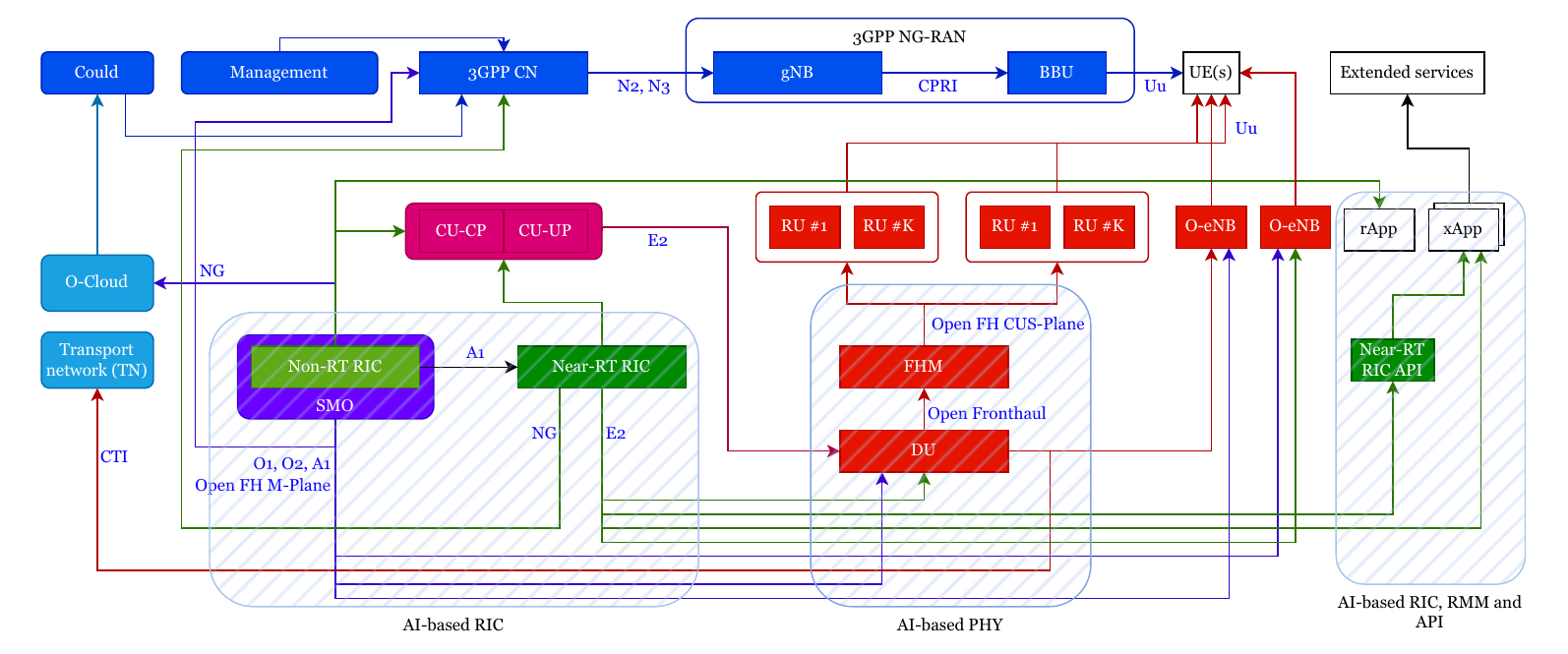}
	\caption{Key elements and interfaces of 3GPP NG-RAN and O-RAN, including the proposed AI-based modules}
	\label{fig_ran_architecture}
\end{figure*}

\subsubsection{Open-RAN Extensions}
O-RAN builds on the 3GPP baseline, adopting Split~7-2x for the Open Fronthaul and introducing additional interfaces, including A1 (policy interface), E2 (near-real-time control interface), O1 (management interface), O2 (cloud management interface), and R1 (xApp–RIC interface), to enhance interoperability~\cite{etsipas103859}. Its main principles are cloudification, open interfaces, and programmability, operationalized via the O-RU (O-RAN Radio Unit), O-DU (O-RAN Distributed Unit), O-CU (O-RAN Central Unit), and the RICs (xApps/rApps) together with the SMO framework.  

\subsubsection{AI-RAN Integration}
AI-RAN initiatives \cite{airan} extend 3GPP/O-RAN by defining AI-native frameworks, including standardized data pipelines, model lifecycle management, and distributed training and inference across device–edge–cloud resources. Reference workloads, such as channel prediction and mobility forecasting, showcase closed-loop optimization in both terrestrial and NTN settings.

\begin{figure*}
	\centering
	\includegraphics[width=.75\linewidth]{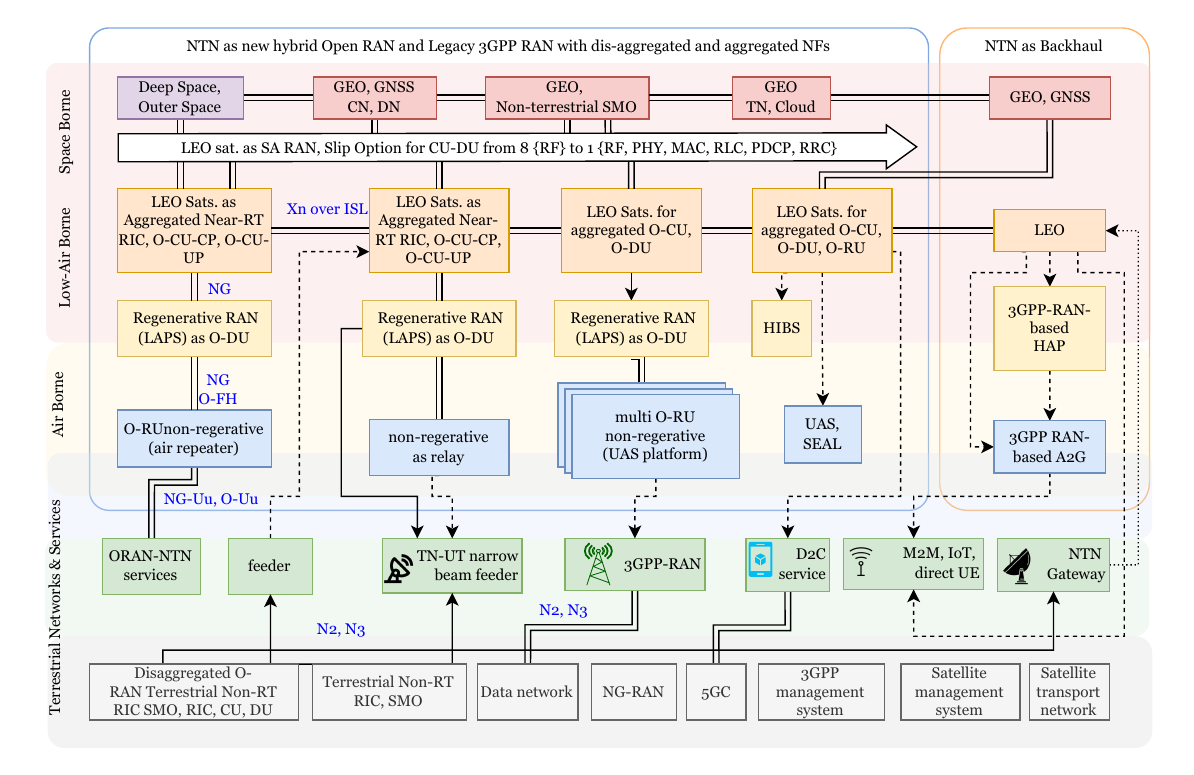}
	\caption{Proposed blueprint of an integrated AIO-RAN-NTN architecture focusing on aggregated NFs and air interfaces}
	\label{fig_ntn_interfaces}
\end{figure*}

\vspace{-.5em}
\subsection{Key Interfaces for O-RAN and AI-RAN in NTN}

To realize AIO-RAN-NTN, several interfaces require adaptation for NTN environments, as shown in Fig.~\ref{fig_ntn_interfaces}.  
At the NG-RAN–core boundary, the N2 (NG Control, NG-C) and N3 (NG User, NG-U) interfaces connect the gNB-CU control and user planes, respectively, to the 5G Core (5GC) functions Access and Mobility Management Function (AMF) and User Plane Function (UPF). In O-RAN, these retain the same 3GPP protocol stacks, mapped to the O-CU-CP and O-CU-UP.  

Within the RAN, the A1 and E2 interfaces enable programmability and AI integration. A1 connects the Non-RT RIC (in the SMO) with the Near-RT RIC to exchange policies and guidance, while E2 connects the Near-RT RIC to O-CU and O-DU elements, supporting real-time control via the E2 Application Protocol and service models.  

The Open Fronthaul (FH) interface connects O-DU and O-RU nodes, supporting split options 7 and 8 at the Physical (PHY) layer. It comprises the Control, User, and Synchronization (CUS) planes and the Management (M) plane, with the latter also linking O-RUs to the SMO.  

For the user plane, the Uu (NG-Uu) interface connects User Equipment (UE) to the gNB across the full NR protocol stack. Although not redefined by O-RAN, it remains central in NTN scenarios where latency, Doppler, and handover dynamics must be considered. O-RAN’s modularity also allows some internal interfaces (e.g., F1-C/U, E1, E2) to be collapsed in deployment, highlighting its scalability for NTN environments.

\subsection{Blueprint for Integrated AIO-RAN-NTN: Proposed Architecture with Open Interfaces}

3GPP Rel-17 and Rel-18 introduced NR-based NTN capabilities, including PHY-layer adaptations and regenerative payloads on Low Earth Orbit (LEO) satellites capable of hosting gNB functions. Building on these foundations, we propose an integrated AIO-RAN-NTN architecture that unifies terrestrial and non-terrestrial access while exposing traditionally internal RAN interfaces as open, NTN-capable air/space links. The design targets a fully Standalone (SA) 5G deployment, with Non-Standalone (NSA) interoperability preserved during the transition phase. A central challenge in NTNs remains the trade-off between mobility and delay, requiring differentiated handling of delay-tolerant services and mobility-sensitive applications.

\subsubsection{System Blueprint}
\emph{Management/Orchestration (SMO/O-Cloud):}
O-RAN SMO spans ground edge and cloud. O1 manages Network Function (NF) lifecycle, O2 manages O-Cloud resources, A1 steers policies to the Non-RT RIC, and E2 controls and observes Near-RT RIC/xApps across TN and NTN.  
\emph{AI-RAN Layer:}
The AI-RAN framework overlays this blueprint, providing standardized data pipelines, feature stores, and lifecycle management. Models for mobility prediction, Doppler compensation, and beam-hopping optimization are trained across distributed TN and NTN data and orchestrated by the Non-RT RIC, with inference hosted in Near-RT RIC/xApps and device/edge accelerators. This ensures closed-loop optimization across terrestrial and non-terrestrial domains.

\subsubsection{New Elements Needed in Our Blueprint}

We propose the following new elements to enable the blueprint:  

\begin{itemize}
	
	\item {A-OFH:} The NTN Aerial Open Fronthaul Profile extends the existing OFH with robustness modes, adaptive numerology, Doppler-aware tracking, and synchronization-quality metrics.  
	
	\item {NTN-IAB:} Open NR-Integrated Access and Backhaul (IAB) extensions are adapted for space and aerial donors or relays. They include support for beam-hopping and contact-window management.  
	
	\item {Delay-Aware E2/xApps:} These extensions incorporate prediction horizons and actuation deadlines. AI-enabled xApps support tasks such as pass prediction, Hybrid Automatic Repeat Request (HARQ) budget control, and channel robustness.  
	
	\item {Federated rApps/xApps:} Federated learning (FL) is applied across distributed terrestrial network (TN) and NTN domains~\cite{Do2023}. The Non-RT RIC orchestrates training, while aggregation occurs at Near-RT RICs for enhanced mobility and blockage prediction.  
	
	\item {Energy- and Cost-Aware NF Placement:} AI-guided placement of RU, DU, and CU-UP functions is carried out across ground and aerial resources, subject to power, budget, and Key Performance Indicator (KPI) constraints.  
	
\end{itemize}

\subsubsection{Element Blocks and Interfaces}

Enabled by the proposed elements, we define three blueprint options:  

\begin{itemize}[leftmargin=.9cm]
	
	\item[\textbf{BP-1}] {Disaggregated NTN-RAN (aerial RU with DU/CU on the ground):}  
	This option uses the A-OFH profile (7.2x) over NTN and the F1 interface (Option~2) over a delay-tolerant backhaul. NTN adaptations include Doppler and timing advance compensation, extended protocol timers, jitter buffers, and holdover synchronization with sync-quality monitoring. AI-based Radio Resource Management (RRM) is supported through the Near-RT RIC over the E2 interface, with delay-aware schedulers, beam-hopping awareness, AI-driven channel forecasting, HARQ budget control, and pass prediction.  
	
	\item[\textbf{BP-2}] {Aggregated NTN-RAN (aerial RU+DU with CU on the ground, or aerial RU+DU+CU-UP):}  
	This option collapses the interfaces defined in BP-1 to reduce latency and signaling overhead. Fronthaul bandwidth requirements are further reduced through the use of NTN-IAB. The system employs minimal A-OFH together with NTN-based F1, N2, and N3 interfaces, reinforced by sync-quality metrics to maintain reliable performance. Energy-aware duty cycling is applied on aerial and space platforms, guided by AI-driven workload placement and federated rApps/xApps for distributed mobility and blockage prediction.  
	
	\item[\textbf{BP-3}] {Hybrid NSA to SA Migration (with legacy 3GPP support):}  
	This option extends BP-1 and BP-2. NTN-IAB is used for TN/NTN traffic steering and splitting, combined with energy- and cost-aware NF placement. Federated rApps/xApps enable cross-domain learning between TN and NTN to improve mobility and blockage resilience. The RIC topology includes a centralized Non-RT RIC and distributed Near-RT RIC instances deployed at the edge or on aerial platforms to reduce control-loop delay.  
	
\end{itemize}

\section{Demo: AI-based KPI Forecasting-empowered OpenAirInterface-based SA Architecture}

\subsection{OpenAirInterface Platform}

OpenAirInterface (OAI) is an open-source project that implements 3GPP-compliant cellular technologies on general-purpose x86/Linux platforms with software-defined radio hardware such as USRP devices~\cite{oai}. The current OAI 5G RAN project develops both Non-Standalone and Standalone gNB implementations, as well as NSA/SA UEs.  
OAI has recently been extended with O-RAN components. In particular, the FlexRIC platform provides a flexible, O-RAN–compliant RAN Intelligent Controller (RIC) and an E2 agent integrated with the OAI radio stack~\cite{oai}. This enables interoperability with O-RAN E2 interfaces and allows control through xApps, such as for network slicing in NSA 5G deployments.

\subsection{Experiment Setup and Downlink Test}

\begin{figure}[htp]
    \centering
    \subfloat[]{%
	\includegraphics[width=.155\linewidth]{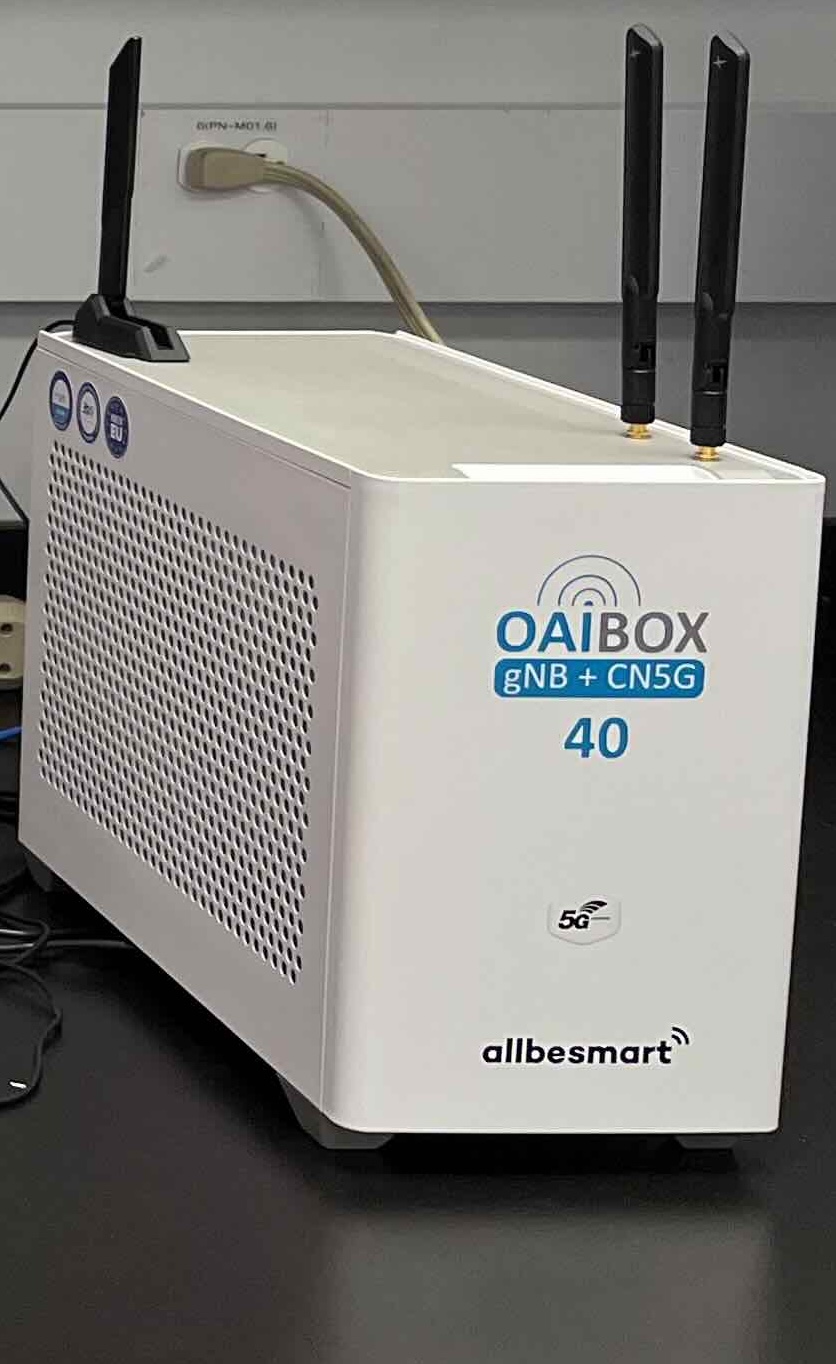}
		\label{fig_bs}
	}
    \hfil
    \subfloat[]{%
		\includegraphics[width=0.15
		\linewidth]{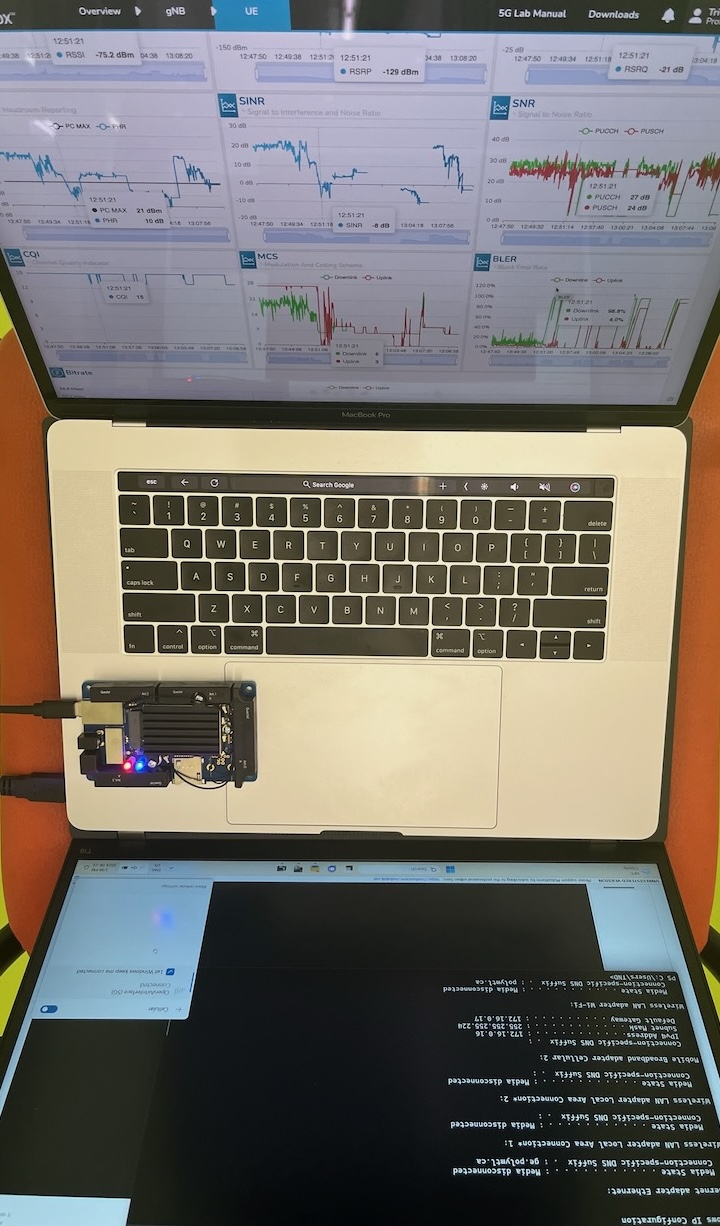}
		\label{fig_mobile_ue}
	}
    \hfil
    \subfloat[]{%
		\includegraphics[width=0.15
		\linewidth]{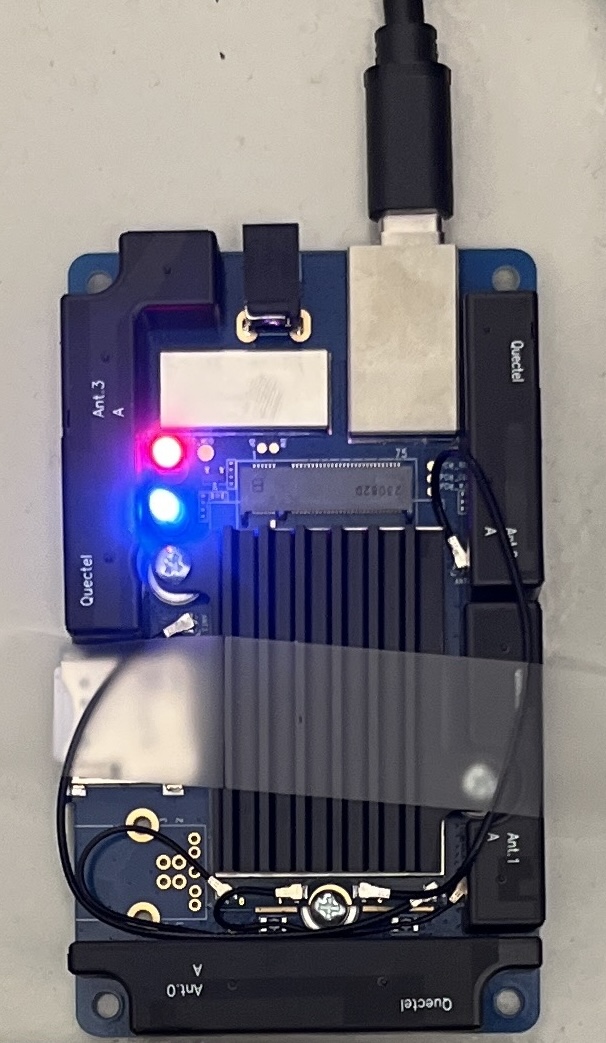}
		\label{fig_fixed_ue}
	}
    \hfil
    \subfloat[]{%
    \includegraphics[width=0.25\linewidth]{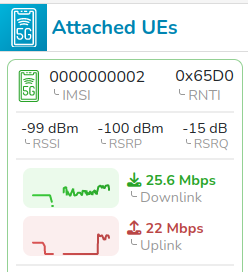}
		\label{fig_bs}
	}
\caption{(a) OAI-gNB, (b) mobile UE, (c) UE Quectel, and (d) UE dashboard}
    \label{fig_testbed}
\end{figure}

We demonstrate an over-the-air transmission using a 3GPP-standard-compliant Standalone 5G system implemented on OAI. The setup consists of a complete end-to-end configuration including:  $(i)$ a 5G Core Network,  $(ii)$ an OAI-based 5G gNodeB, and  $(iii)$ mobile user equipment, as shown in Fig.~\ref{fig_testbed}.
This testbed allows us to analyze mobility-induced performance degradation and collecting data for AI-based KPI forecasting.

\begin{figure}
    \centering
    \includegraphics[width=1\linewidth]{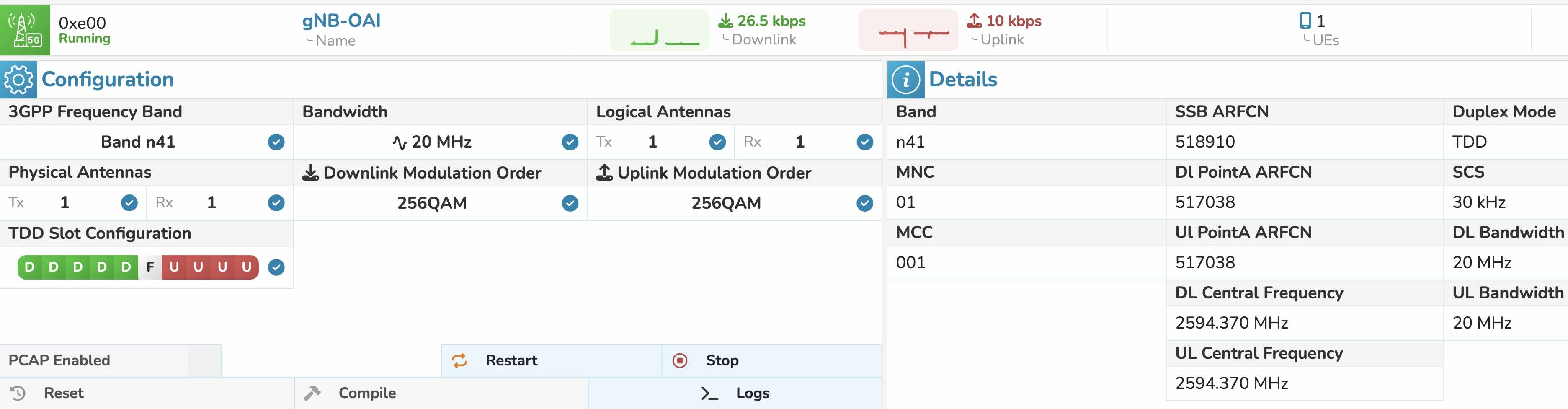}
\caption{Hardware and signal processing setup at gNB.}
    \label{fig_bs_config}
\end{figure}

\subsubsection{gNB} 
The gNB is implemented on an OAIBOX-40 \cite{oaibox} running the open-source OAI 5G stack, with CU and DU functions integrated on a single host alongside the 5G Core. The RU is realized with a software-defined radio (SDR) supporting up to 40~MHz bandwidth.  

\subsubsection{UE}  
The UE is a Quectel 5G module with integrated antennas, as shown in 
Figs.~\ref{fig_mobile_ue}--\ref{fig_fixed_ue}. It is operated via a programmed 
SIM card and controlled from a laptop using Quectel drivers and a terminal 
interface (MobaXterm).

\subsubsection{Configuration and Downlink Test}  
The gNB parameters (e.g., frequency, time structure, modulation/coding scheme) 
are configured online through a web-based GUI dashboard, as shown in our repository. Downlink performance is evaluated using iPerf, with 
the gNB acting as the iPerf server and transmitting UDP traffic to the UE 
terminal. To capture a variety of propagation scenarios, the UE is moved to 
experience both LOS and NLOS channel conditions.  
For monitoring purposes, end-to-end KPI results are collected in real time 
(with one-second resolution) and stored in JSON format.

\begin{figure}[htp]
\centering
    \subfloat[]{%
    \includegraphics[width=1\linewidth]{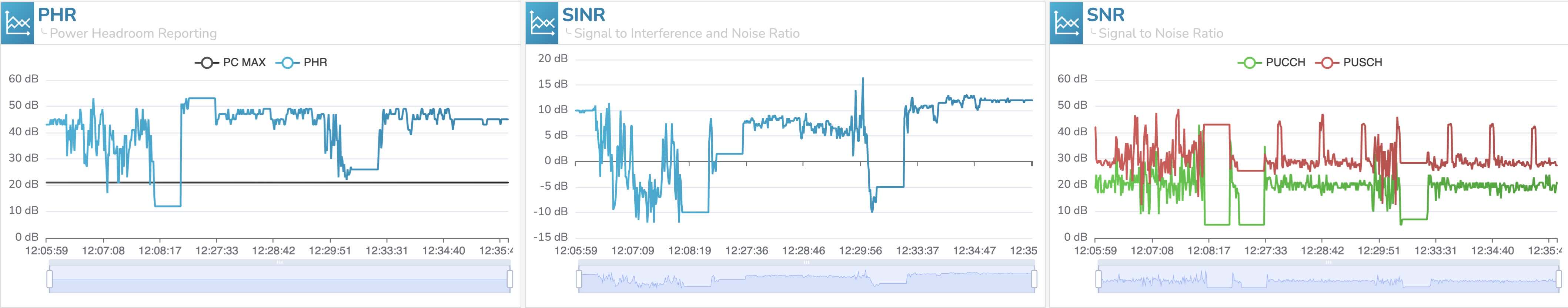}
    \label{fig_kpi_b}
    }
    \\
    \subfloat[]{%
    \includegraphics[width=0.65\linewidth]{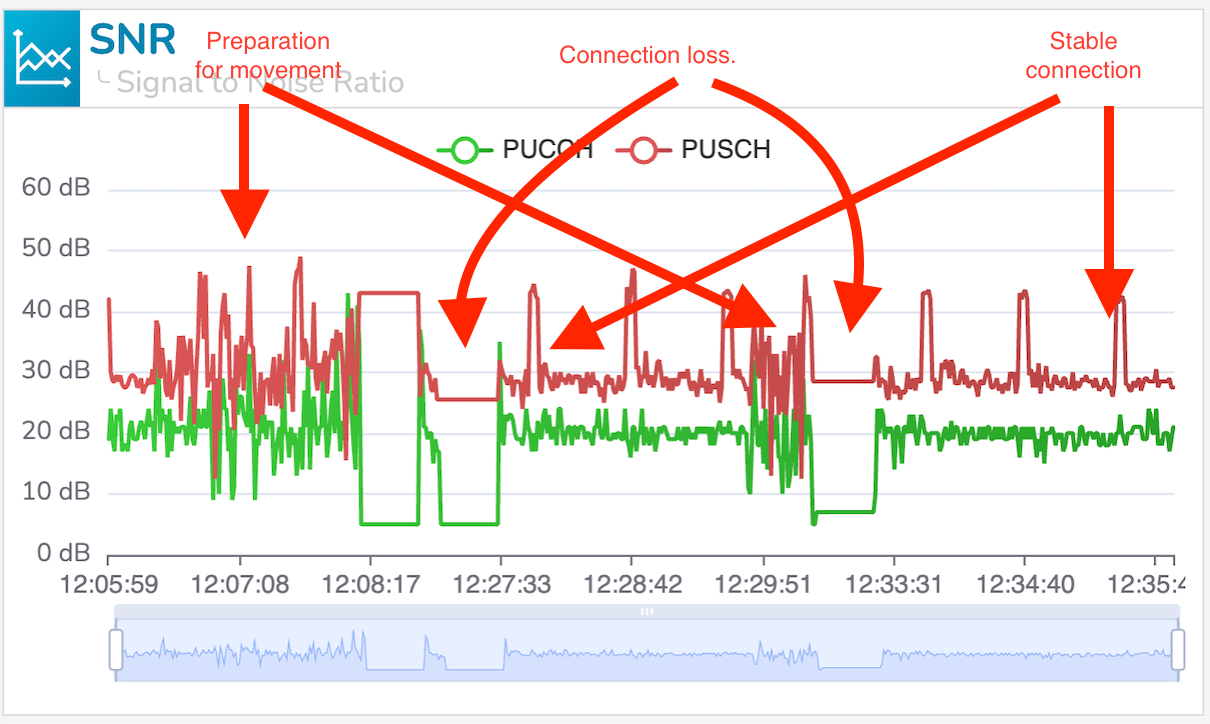}
    \label{fig_snr_observation}
    }
\caption{Key observed KPIs with comments on connectivity}
\label{fig_kpi_all}
\end{figure}

\subsection{Time Series Analysis and LSTM-based KPI Forecasting}

As shown in Fig.~\ref{fig_kpi_all}, key performance indicators such as RSSI, SINR, SNR, PHR, etc. fluctuate significantly with UE mobility. These variations force the gNB to continuously adapt the modulation and coding scheme (MCS), and frequent drops may cause connection loss. In our OAI-based setup, reconnection requires manual intervention with noticeable delay, underscoring the need for proactive forecasting of SINR and related features.  

\vspace{-1em}
\subsection{Feature Extraction: Extract from JSON Data}

We represent the dataset as a JSON document $D$ drawn from the class of rooted,
finite, labeled trees $\mathcal{T}$ whose internal nodes are dictionaries
(key--value maps) and whose leaves are primitives (numbers, strings, booleans),
lists, or dictionaries, i.e., $D \in \mathcal{T}$.

\paragraph{Loading Raw Data}
Given a recording file $p$, we define the (partial) loading operator $\mathsf{Load}(p)$. This operator allows us to parse the JSON file $p$ so that we can extract and return the corresponding dataset $D$.

\paragraph{Key-based Extraction}  
For a given key $k$ (a string), we define the extraction operator
$
\mathsf{E}_k : \mathcal{T} \to \mathcal{L},
$
which we use to traverse the dataset $D$ and collect \emph{all} values directly
associated with $k$. We perform this traversal in depth-first, encounter order,
so that
$
\mathsf{E}_k(D) = \bigl(v_1, v_2, \dots, v_m\bigr),
$
where each $v_i$ is either a subtree or a primitive found at some occurrence of $k$.

\paragraph{Two-layer Feature Extraction}  
We fix the keys corresponding to the parameters of interest, e.g., $k_1=\texttt{ues}$ and $k_2 \in \{\texttt{sinr}\}$.  
First, we apply the extraction operator to obtain
$
	\mathcal{S} = \mathsf{E}_{k_1}(D) = (S_1, \dots, S_m), S_i \in \mathcal{T}.
$
Next, for each sub-tree $S_i$, we extract the values associated with $k_2$, giving
$
	\vec{y}_i = \mathsf{E}_{k_2}(S_i) 
	= \bigl(y_{i1}, \dots, y_{i\ell_i}\bigr).
$
Finally, by collecting all such vectors, we obtain
$
\mathcal{Y} = (\vec{y}_1, \dots, \vec{y}_m).
$

\paragraph{Tabularization}  
Let $L = \max_i \ell_i$, and define the padding map
$
\mathsf{Pad} : \mathcal{L} \to \mathbb{R}^L \cup \{\text{NaN}\}^L,
$
which we use to right-pad shorter sequences with \textsc{NaN}s.  
By stacking the padded rows, we obtain
\begin{align}
	Y =
	[
		\mathsf{Pad}(\vec{y}_1)^\top
		\ldots
		\mathsf{Pad}(\vec{y}_m)^\top
	]^\top
	\in (\mathbb{R} \cup \{\text{NaN}\})^{m \times L}.
\end{align}
We then regard the resulting matrix $Y$ as a multivariate time series,
$
	\mathcal{D} = \{x_t \in \mathbb{R}^d\}_{t=1}^N,
$
with $N$ time steps and $d$ features (columns).  
Finally, we export $Y$ to CSV format so that we can feed it into our neural network model.

\subsection{Feature Engineering: Construct Time-Series Data}

\paragraph{Imputation of Disconnections}  
Let $F_i(t)$ denote feature $i$ at time $t$. If the UE is disconnected at $t$,  
then we set $F_i(t) = \texttt{NaN}$. To handle such cases, we replace undefined entries 
by a constant $\kappa$, which is chosen heuristically based on the training experience. Thus, we have
\begin{align} \label{eq_disconnection}
	F_i(t) \leftarrow
	\begin{cases}
		\kappa, & \text{if the UE is disconnected at } t, \\
		F_i(t), & \text{otherwise},
	\end{cases}
\end{align}
where $\kappa \propto \mathbb{E}[F_i(t) \mid \text{UE connected}].$

\emph{Remark.} It is noted that $\kappa$ plays a significant role during training, 
as it encodes the separation between connected and disconnected regimes.

\paragraph{Data Split}
Let $N = |\mathcal{D}|$. We define disjoint, contiguous partitions
$\mathcal{D}_{\mathrm{train}} = \{x_t\}_{t=1}^{\lfloor 0.7N \rfloor}$,
$\mathcal{D}_{\mathrm{val}} = \{x_t\}_{t=\lfloor 0.7N \rfloor + 1}^{\lfloor 0.9N \rfloor}$, and
$\mathcal{D}_{\mathrm{test}} = \{x_t\}_{t=\lfloor 0.9N \rfloor + 1}^{N}$,
where $d = \dim(\mathcal{D})$ denotes the number of features.

\paragraph{Scaling $\min-\max$}
To avoid leakage, we fit the feature-wise min–max scaling on the \emph{training} split only.
For each feature $j\in\{1,\dots,d\}$,
$
x_{\min}^{(j)}=\min_{x\in \mathcal{D}_{\mathrm{train}}} x^{(j)},\qquad
x_{\max}^{(j)}=\max_{x\in \mathcal{D}_{\mathrm{train}}} x^{(j)}.
$
Let $S : \mathbb{R}^d \to [0,1]^d$ be the scaling operator defined as
\begin{align} \label{eq_scaler}
	\bigl(S(x)\bigr)^{(j)} =
	(x^{(j)} - x_{\min}^{(j)})/(\,x_{\max}^{(j)} - x_{\min}^{(j)}\,),
\end{align}
with $x_{\min}^{(j)}$ and $x_{\max}^{(j)}$ taken from the training data.
If $x_{\max}^{(j)} = x_{\min}^{(j)}$, set $(S(x))^{(j)}=0$.
Applying $S$ to each split yields
\begin{align}
\tilde{\mathcal{D}}_{\mathrm{train}}=S(\mathcal{D}_{\mathrm{train}}),
\tilde{\mathcal{D}}_{\mathrm{val}}=S(\mathcal{D}_{\mathrm{val}}),
\tilde{\mathcal{D}}_{\mathrm{test}}=S(\mathcal{D}_{\mathrm{test}}).
\end{align}

\paragraph{Windowing with Shift and Labels}
Let the window parameters be
$w_{\mathrm{in}} \ (\text{input width})$  
$w_{\mathrm{lbl}} \ (\text{label width})$  
$s \ (\text{shift})$  
$w_{\mathrm{tot}} = w_{\mathrm{in}} + s$.
For a split with length $N_{\mathtt{set}}$ ($\mathtt{set}\in\{\mathrm{train,val,test}\}$) and
$t\in\{1,\dots,N_{\mathtt{set}}-w_{\mathrm{tot}}-w_{\mathrm{lbl}}+1\}$, define
\begin{align*}
X_t &= \tilde{\mathcal{D}}_{\mathtt{set}}[t : t + w_{\mathrm{in}} - 1, :]
    \in [0,1]^{w_{\mathrm{in}} \times d}, \\
Y_t &= \tilde{\mathcal{D}}_{\mathtt{set}}[t + w_{\mathrm{tot}} - w_{\mathrm{lbl}} : 
                                  t + w_{\mathrm{tot}} - 1, :]
    \in [0,1]^{w_{\mathrm{lbl}} \times d'}.
\end{align*}
By sliding the window, we obtain
\begin{align}
\mathcal{W}_{\mathtt{set}}=\{(X_t,Y_t)\}_{t=1}^{M_{\mathtt{set}}},
M_{\mathtt{set}} \!=\! N_{\mathtt{set}} - w_{\mathrm{tot}} - w_{\mathrm{lbl}} + 1.
\end{align}

\subsection{Mathematical Model of LSTM Training}

\paragraph{Composite Architecture}
Let $T = \texttt{time\_steps} = w_{\mathrm{in}}$ and $d = \texttt{input\_dim}$, 
as specified in the source code. For an input sequence 
$X \in \mathbb{R}^{T \times d}$, the model computes
$
\hat{Y} = f_{\mathrm{NN}}(X) \in \mathbb{R}^{T \times C},
$
where
$
	f_{\mathrm{NN}} = f_{\mathrm{out}} \circ \cdots \circ f_{\mathrm{fci}} \circ \cdots
	\circ f_{\mathrm{LSTMj}} \circ \cdots
$
denotes the composition of layers. Specifically, the network consists of 
$N_{\mathrm{LSTM}}$ stacked LSTM layers, followed by 
$N_{\mathrm{FC}}$ fully connected (time-distributed) layers with ReLU activation, 
and a final linear output layer of width $C$. 
Further details on the LSTM architecture can be found in the source code.

\paragraph{Label Alignment}
If $w_{\mathrm{lbl}}\le T$, let $\Pi_{\mathrm{lbl}}:\mathbb{R}^{T\times C}\to
\mathbb{R}^{w_{\mathrm{lbl}}\times C}$ select the final $w_{\mathrm{lbl}}$ time
indices:
$
	\hat{Y}^{(\mathrm{lbl})}=\Pi_{\mathrm{lbl}}(\hat{Y}).
$
Training targets are given by 
$
Y \in [0,1]^{w_{\mathrm{lbl}} \times d'},
$
which are taken from the window labels, where $d'$ denotes the dimension 
of the selected label features. If $C \neq d'$, we apply a fixed 
feature-selector mapping
$
Y \mapsto Y' \in [0,1]^{w_{\mathrm{lbl}} \times C}
$
to ensure compatibility with the model output width $C$. 
It is noted that, in our time-series data, the labels correspond to the 
data itself observed at different sampling instances.

\paragraph{Loss in Scaled Domain}  
With batches indexed by $i$ and time by $\tau$, the loss function is defined as  
\begin{align}
	\mathcal{L}_{\mathrm{MSE}}
	= \frac{1}{B\,w_{\mathrm{lbl}}\,C}
	\sum_{i=1}^{B}\sum_{\tau=1}^{w_{\mathrm{lbl}}}\sum_{c=1}^{C}
	\bigl(\hat{Y}^{(\mathrm{lbl})}_{i,\tau,c} - Y_{i,\tau,c}\bigr)^2,
\end{align}
where $B$ is the batch size, $w_{\mathrm{lbl}}$ the label window length, 
and $C$ the output dimension.  
Model parameters are collected in
$
\Theta = \{ W^{(1:N_{\rm LSTM})},\, b^{(1:N_{\rm LSTM})},\, \theta_{\mathrm{LSTM}} \},
$
and are updated using a chosen optimizer, with learning rate $\eta$, as
\begin{align}
	\Theta \leftarrow \Theta - \eta \,\nabla_{\Theta}\mathcal{L}_{\mathrm{MSE}},
\end{align}

\paragraph{Training Phase}  
It is noted that each window $X_t \in [0,1]^{w_{\mathrm{in}} \times d}$ is fed as a 
sequence of length $T = w_{\mathrm{in}}$. The model then produces 
$\hat{Y}_t \in \mathbb{R}^{T \times C}$, which is projected to the label slice 
$\hat{Y}^{(\mathrm{lbl})}_t$ and compared to $Y_t$.  
The training procedure is summarized in Algorithm~\ref{alg}.

\begin{algorithm}[htp]
	\label{alg}
	\caption{Proposed end-to-end pipeline for KPI prediction: load $\to$ extract $\to$ impute $\to$ split $\to$ scale $\to$ window $\to$ train $\to$ test $\to$ inverse-scale $\to$ RMSE}
	\DontPrintSemicolon
	\KwIn{JSON path $p$; keys $k_1=\texttt{ues}$, $k_2\in\{\texttt{sinr}\}$; imputation constant $\kappa$; split ratios $(0.7,0.2,0.1)$; window params $(w_{\mathrm{in}},w_{\mathrm{lbl}},s)$; model hparams $(u,h,C,\eta)$; epochs $E$.}
	\KwOut{Trained parameters $\Theta$; $\mathrm{RMSE}_{\mathrm{tr}}$, $\mathrm{RMSE}_{\mathrm{te}}$ (original units).}
	
	\textbf{Load \& extract}\;
	$D \leftarrow \mathsf{Load}(p)$ 
	$\mathcal{S} \leftarrow \mathsf{E}_{k_1}(D)=(S_1,\dots,S_m)$ 
	$\mathcal{Y} \leftarrow (\mathsf{E}_{k_2}(S_1),\dots,\mathsf{E}_{k_2}(S_m))$ 
	$Y \leftarrow \textsf{PadStack}(\mathcal{Y}) \in (\mathbb{R}\cup\{\text{NaN}\})^{m\times L}$ \;
	
	\textbf{Impute}\;
	\ForEach{entry $y$ in $Y$}{
		\If{$y=\text{NaN}$}{ $y \leftarrow \kappa$ }
	}
	Form time series $\mathcal{D}=\{x_t\in\mathbb{R}^d\}_{t=1}^{N}$ from $Y$\;
	
	\textbf{Split}\;
	$N_{\mathrm{tr}}=\lfloor 0.7N\rfloor$, $N_{\mathrm{val}}=\lfloor 0.2N\rfloor$, $N_{\mathrm{te}}=N-N_{\mathrm{tr}}-N_{\mathrm{val}}$\;
	$\mathcal{D}_{\mathrm{train}},\ \mathcal{D}_{\mathrm{val}},\ \mathcal{D}_{\mathrm{test}} \leftarrow$ contiguous partitions of $\mathcal{D}$\;
	
	\textbf{Fit scaler (train only)}\;
	For each feature $j$: $x_{\min}^{(j)}=\min_{x\in\mathcal{D}_{\mathrm{train}}}x^{(j)}$, $x_{\max}^{(j)}=\max_{x\in\mathcal{D}_{\mathrm{train}}}x^{(j)}$\;
	Define $S$ by $\bigl(S(x)\bigr)^{(j)}=\frac{x^{(j)}-x_{\min}^{(j)}}{x_{\max}^{(j)}-x_{\min}^{(j)}}$\;
	
	\textbf{Transform splits}\;
	$\tilde{\mathcal{D}}_{\mathrm{train}}=S(\mathcal{D}_{\mathrm{train}})$,\quad
	$\tilde{\mathcal{D}}_{\mathrm{val}}=S(\mathcal{D}_{\mathrm{val}})$,\quad
	$\tilde{\mathcal{D}}_{\mathrm{test}}=S(\mathcal{D}_{\mathrm{test}})$\;
	
	\textbf{Window}\;
	$w_{\mathrm{tot}}=w_{\mathrm{in}}+s$\;
	\For{$t=1$ \KwTo $N_{\mathtt{set}}-w_{\mathrm{tot}}-w_{\mathrm{lbl}}+1$ for each $\mathtt{set}\in\{\mathrm{train,val,test}\}$}{
		$X_t=\tilde{\mathcal{D}}_{\mathtt{set}}[t:t+w_{\mathrm{in}}-1,:]$,\quad
		$Y_t=\tilde{\mathcal{D}}_{\mathtt{set}}[t+w_{\mathrm{tot}}-w_{\mathrm{lbl}}:t+w_{\mathrm{tot}}-1,:]$\;
	}
	$\mathcal{W}_{\mathtt{set}}=\{(X_t,Y_t)\}$ and batch into datasets\;
	
	\textbf{Initialize model}\;
	Set $T=w_{\mathrm{in}}$, $d=\dim(\mathcal{D})$ and build $f_{\Theta}$:
	$\text{LSTM}(u)\times3 \rightarrow \text{FC}(4h)\rightarrow \text{FC}(2h)\rightarrow \text{FC}(h)\rightarrow \text{Linear}(C)$\;
	
	\textbf{Phase: Train}\;
	\For{$e=1$ \KwTo $E$}{
		Update $\Theta \leftarrow \Theta - \eta \nabla_{\Theta}\mathcal{L}_{\mathrm{MSE}}$ on $\mathcal{W}_{\mathrm{train}}$ (validate on $\mathcal{W}_{\mathrm{val}}$)\;
	}
	
	\textbf{Phase: Test \& inverse-scale}\;
	Compute $\hat{Y}^{\mathrm{tr}}=f_{\Theta}(X^{\mathrm{tr}})$,\quad $\hat{Y}^{\mathrm{te}}=f_{\Theta}(X^{\mathrm{te}})$\;
	$\widehat{Y}^{\,\mathrm{tr}}_{\mathrm{orig}}=S^{-1}(\hat{Y}^{\mathrm{tr}})$,\quad
	$\widehat{Y}^{\,\mathrm{te}}_{\mathrm{orig}}=S^{-1}(\hat{Y}^{\mathrm{te}})$,\quad
	$Y^{\mathrm{tr}}_{\mathrm{orig}}=S^{-1}(Y^{\mathrm{tr}})$,\quad
	$Y^{\mathrm{te}}_{\mathrm{orig}}=S^{-1}(Y^{\mathrm{te}})$\;
	
	\textbf{RMSE}\;
	$\mathrm{RMSE}_{\mathrm{phase}}=\sqrt{\frac{1}{N_{\mathrm{phase}}}\sum_{i}(\widehat{Y}^{\,\mathrm{phase}}_{\mathrm{orig},i}-Y^{\mathrm{phase}}_{\mathrm{orig},i})^2}$\;
\end{algorithm}

\paragraph{Testing and Inverse Scaling}  
During evaluation, we draw a batch 
$(X^{\mathrm{test}}, Y^{\mathrm{test}}) \sim \mathcal{W}_{\mathrm{test}}$ 
and compute scaled predictions 
$
\hat{Y}^{\mathrm{test}} = f_{\Theta}(X^{\mathrm{test}}).
$
Let $S^{-1}$ denote the inverse of the train-fitted scaler $S$, applied 
componentwise. Predictions and labels are then reported in the original units:
$
\widehat{Y}^{\,\mathrm{test}}_{\mathrm{orig}} = 
S^{-1}\!\bigl(\hat{Y}^{\mathrm{test}}\bigr), 
Y^{\mathrm{test}}_{\mathrm{orig}} = 
S^{-1}\!\bigl(Y^{\mathrm{test}}\bigr).
$
Flattening across batch, time, and feature dimensions yields 
a total sample count of $N_{\mathrm{tr}}$ for training and $N_{\mathrm{te}}$ for 
testing. Performance is then evaluated in terms of RMSE as:
\begin{align}
	\mathrm{RMSE}_{\mathrm{tr}}
	= \bigg[\frac{1}{N_{\mathrm{tr}}}\sum_{i=1}^{N_{\mathrm{tr}}}
		\bigl(\widehat{Y}^{\,\mathrm{tr}}_{\mathrm{orig},i}
		- Y^{\mathrm{tr}}_{\mathrm{orig},i}\bigr)^2\bigg]^{1/2}.
\end{align}


\section{Results and Discussion}

Following Alg.~\ref{alg}, the LSTM models can be trained either on the host 
laptop (with over-the-air feedback) or directly on the OAIBOX running a 
Linux platform. Due to space limitations, all specific settings and datasets 
are provided in our GitHub repository referenced above.
In our training, the loss function converges as shown in our code repository.
It is noted that, due to the nature of the raw data, specifically the disconnected events between the gNB and UE, we apply the feature-wise scaler $S$ in \eqref{eq_scaler} to normalize values into the range $[0,1]$. The constant $\kappa$ is set near the lower bound (typically $\approx 0$, or slightly below). In this way, scaling preserves the contrast between connected and disconnected regimes while placing all inputs on a comparable numerical scale, which stabilizes LSTM training and accelerates convergence.

\begin{figure}
    \centering
    \includegraphics[width=.85\linewidth]{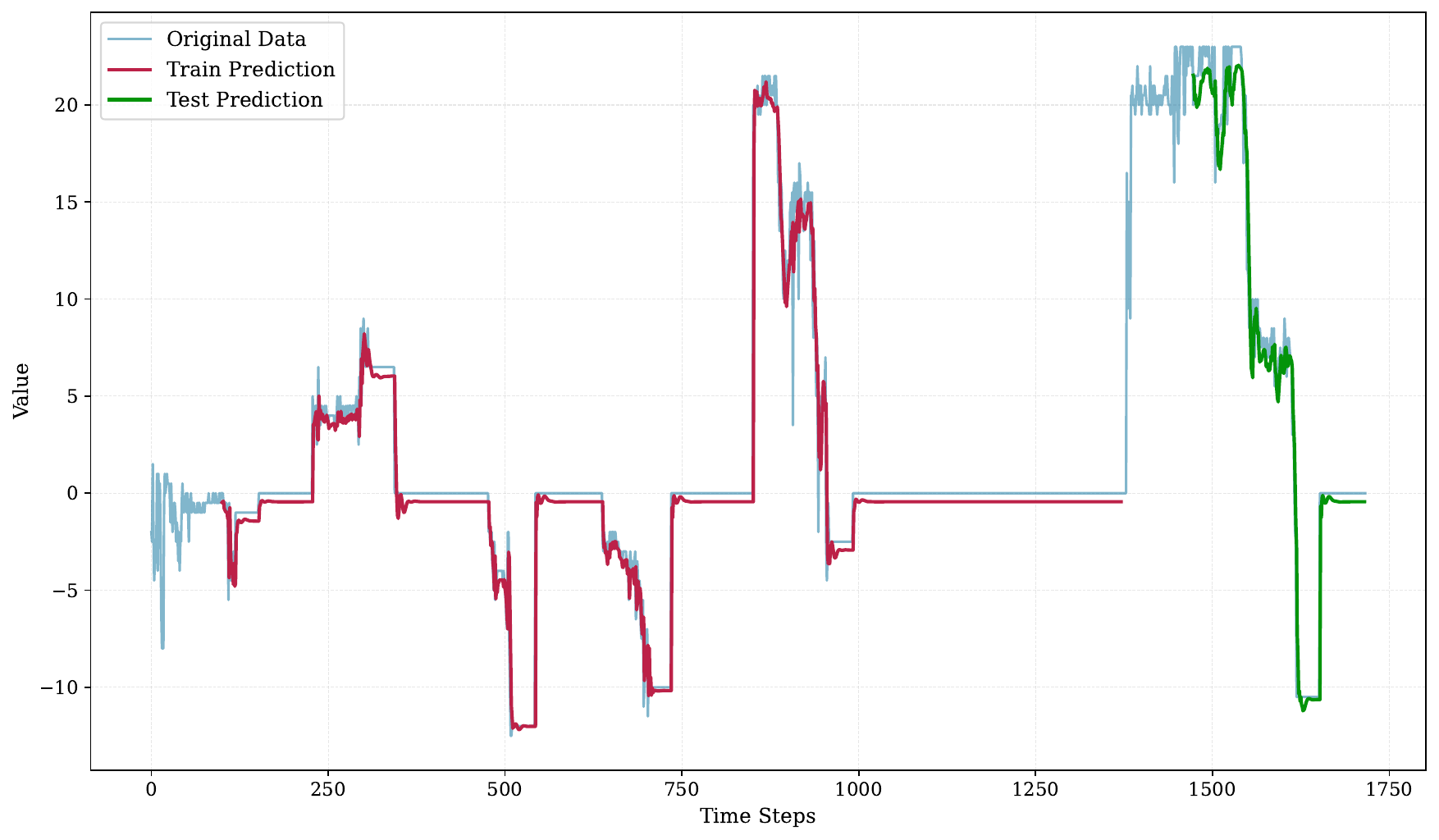}
\caption{Predicted KPI, i.e., SINR with $\kappa=0$}
    \label{fig_sinr_predict}
\end{figure}
\begin{figure}[htp]
	\centering
	\includegraphics[width=\linewidth]{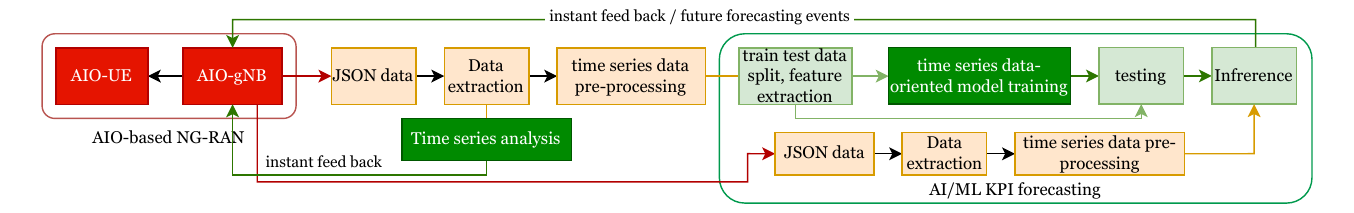}
\caption{Proposed KPI prediction for adaptive operation in AIO-RAN-SA.}
	\label{fig_pipeline}
	\vspace{-1em}
\end{figure}

As shown in Fig.~\ref{fig_sinr_predict}, we are able to successfully predict the 
KPI of interest into the future, albeit with some delay. It is noted that the 
gap observed in Fig.~\ref{fig_sinr_predict} arises from the scaling and the 
inclusion of $\kappa$. Specifically, because disconnections yield missing values, 
the constant $\kappa$ introduced in \eqref{eq_disconnection} is a key factor in 
the data processing pipeline. Its presence creates a pronounced numerical gap 
between the disconnected level $\kappa$ and the connected events.  
Nevertheless, the results are sufficiently accurate for binary forecasting tasks 
such as distinguishing connected versus disconnected states, or monitoring LOS/NLOS 
transitions.

In this work, the forecasting module operates in a monitoring mode only. 
As future work, APIs will be exposed for integrating the prediction module 
directly into the OAI gNB stack, thereby enabling closed-loop adaptation of 
RAN parameters. To this end, we have proposed a pipeline as shown in Fig.~\ref{fig_pipeline}, that combines time-series analysis with 
AI/ML-based forecasting applied to the collected data.

\vspace{-.5em}
\section{Conclusions}

In this paper, we proposed {AIO-RAN-NTN}, a blueprint that redefines 
internal RAN interfaces as open, NTN-capable links, with the key challenge 
being mobility management in NTNs. We implemented a 3GPP-compliant SA 5G 
testbed using the OAI platform.  
Our experimental results confirmed that mobility-induced channel variations 
degrade performance and cause reconnection delays. To address this issue, 
we developed an AI/ML-based forecasting pipeline for key performance indicators (KPIs), capable of anticipating KPI and SINR evolution. 

\vspace{-1em}
\bibliographystyle{IEEEtran}
\bibliography{References}

\end{document}